\documentclass[preprint,12pt]{elsarticle}

\usepackage{graphicx}
\usepackage{amssymb}

\usepackage{lineno}

\usepackage{array}

\usepackage{pifont}
\def\check{\ding{51}}

\usepackage{url}

\journal{Current Opinion in Systems Biology}

\begin{document}

\begin{frontmatter}


\title{Parameter Estimation and Uncertainty Quantification for Systems Biology Models}



\cortext[cor1]{Corresponding author. wish@lanl.gov}
\author{Eshan D. Mitra}
\author{William S. Hlavacek\corref{cor1}}

\address{Theoretical Biology and Biophysics Group, Theoretical Division, Los Alamos National Laboratory, Los Alamos, NM 87545, USA}

\begin{abstract}
Mathematical models can provide quantitative insight into immunoreceptor signaling, but require parameterization and uncertainty quantification before making reliable predictions. We review currently available methods and software tools to address these problems. We consider gradient-based and gradient-free methods for point estimation of parameter values, and methods of profile likelihood, bootstrapping, and Bayesian inference for uncertainty quantification. We consider recent and potential future applications of these methods to systems-level modeling of immune-related phenomena.

\end{abstract}


\end{frontmatter}

\section*{Highlights}

\begin{itemize}
    \item Models of immunoreceptor signaling often contain parameters that must be fit to data.
    \item New tools including PyBioNetFit and AMICI support automated parameter estimation.
    \item Optimization algorithms can be used to obtain point estimates of parameter values.
    \item Parameter estimation can incorporate both quantitative and qualitative data.
    \item Uncertainty quantification assesses confidence in parameter values and model predictions.
\end{itemize}


\section*{Introduction}

Immunoreceptors such as the T cell receptor (TCR) \cite{Brownlie2013}, B cell antigen receptor (BCR) \cite{DalPorto2004}, and high-affinity IgE receptor (Fc$\epsilon$RI) \cite{Rivera2006} serve as initiation points for information processing by extensive cell signaling networks, which have been characterized over decades of experimental work. 

Mathematical models of the dynamics of these networks can enable new quantitative insights into immune cell signaling, but present challenges: each interaction in the signaling network is characterized by one or more rate constants, which are often unknown. A model encompassing even a small subset of the known protein-protein interactions could have tens to hundreds of unknown parameters. Such models raise problems of estimating parameter values, and quantifying uncertainty in parameter estimates and in model predictions. These challenges are compounded by the fact that the state space (i.e., the number of different chemical species present) can grow large, making simulations computationally demanding. Thus, parameterization tools for biological modeling must deal with a high-dimensional search space while minimizing the number of expensive model simulations. Parameterization and uncertainty quantification, which are our focus in this review, are important aspects of analysis of quantitative models. Another aspect, not covered here, is model selection \cite{Lin2018}.

Various software tools such as COPASI \cite{Hoops2006}, Data2Dynamics \cite{Raue2015}, AMICI \cite{Frohlich2019}, and PyBioNetFit \cite{Mitra2019a} make parameterization of detailed models possible without the need for problem-specific code. PyBioNetFit and AMICI are the newest of these tools, and both provide features that are complementary to those available in older tools. 

Here, we review recent advances in methods and tools that address the parameterization problem. We focus on modeling cell signaling in immunity, but note that the same methodology has applications across systems biology. This review serves as a guide for a systems biology modeler: given a particular model and experimental dataset for a cellular process of interest, we discuss our recommended approaches for parameter estimation and uncertainty quantification and the available software implementations. Our discussion touches on several applications to modeling immunoreceptor signaling that have been enabled by recent methodological developments in parameterization. 

\section*{Model formulation}

We assume that a model of interest has been constructed based on known molecular mechanisms of signaling. Traditionally, model structure (i.e., the set of protein-protein interactions and biochemical reactions included in a model) is defined through a hand-crafted approach, but recent tools have been developed to make this process computer-aided \cite{Gyori2017,Todorov2019}. Ideally the model should be specified in a standardized format to enable compatibility with the general-purpose tools discussed in this review. For immunoreceptor signaling models, BioNetGen language (BNGL) \cite{Faeder2009} is often a useful format because it supports rule-based modeling. Rule-based modeling is a preferred approach to describe biomolecular site dynamics, which are often important in receptor signaling systems \cite{Chylek2013}. Another available format is the Systems Biology Markup Language (SBML) \cite{Hucka2003}, which has a wider range of software support. The current SBML standard (Level 3) \cite{Hucka2015} includes a core language, which can be supplemented with extension packages, such as SBML Multi \cite{Zhang2018}, which supports rule-based modeling. However, software support for extensions is more limited. Software is available to convert from BNGL to SBML \cite{Harris2016}, so BNGL models can benefit from SBML-compatible tools. An advantage of using these standardized formats is the availability of databases of published models -- BioModels Database \cite{LeNovere2006} for SBML models and RuleHub (\url{https://github.com/RuleWorld/Rulehub}) for BNGL models. Models in these databases can be used for benchmarking or as starting points for new modeling studies. 

A model of interest is assumed to describe the concentrations or populations of chemical species over time, but could take several formats. It could be a system of nonlinear ODEs that are numerically integrated to generate deterministic time courses for each chemical species. It could also be a stochastic model simulated using Gillespie's direct method \cite{Gillespie2006}, for example, or a network-free method \cite{Suderman2018}, such as that implemented in NFsim \cite{Sneddon2011}. In network-free methods, the system state is tracked in terms of the set of molecules currently present in the system, without enumerating all possible chemical species and reactions (which could be too many to practically enumerate). 

We also assume that experimental data are available, which are related to quantities that are represented in the model (possibly via a measurement model). In the conventional case, the data are quantitative time courses or dose-response curves. An objective function is specified to measure model misfit to experimental data. One common choice is a (weighted) residual sum of squares function, $\sum_i w_i(y_i-\hat{y_i})^2$ where $y_i$ are experimental data, $\hat{y_i}$ are model predictions, and $w_i$ are constants. One choice for $w_i$ is $1/\sigma_i^2$, where $\sigma_i^2$ is the sample variance associated with $y_i$; this formulation is sometimes called the \textit{chi-squared} objective function.  We consider a less conventional objective function below. 

The parameterization problem becomes a problem of minimizing the chosen objective function.

\section*{Parameter estimation through optimization}

Several classes of methods, which have strengths for different types of problems, are available to perform minimization of the objective function.

\subsection*{Gradient-based optimization}

Gradient-based optimization consists of a family of methods that involve computing the gradient of the objective function with respect to the parameters. Such methods can be classified as first-order (using only first derivatives of the objective function with respect to parameters) or second-order (using both first and second derivatives). First-order methods include gradient descent and stochastic gradient descent, with the latter commonly used in machine learning applications. Modelers often prefer second-order methods, which avoid becoming trapped at saddle points by leveraging the curvature information in the second derivatives. A common second-order choice is the  Levenberg-Marquardt algorithm \cite{More1978}, but this algorithm is specialized to objective functions expressed as a sum of squares (i.e., least squares problems). For the more general case, quasi-Newton methods (e.g., L-BFGS-B \cite{Byrd1995}) can be used. The above algorithms are standard, but for systems biology models, computation of the gradient is not always straightforward. Below we discuss four possible approaches.

The \textit{finite difference approximation} is a naive method in which the gradient is estimated by systematically perturbing each parameter by a small amount. This method is simple and can be applied to any model, but is inefficient for models with high-dimensional parameter spaces. Moreover, performance can be negatively affected by inexact gradient information. 

The \textit{forward sensitivity} method is a more sophisticated method for exact gradient computation (reviewed in \cite{Sengupta2014}). The method is limited to ODE models. The method consists of augmenting the original ODE system with additional variables and equations for the derivative of each species concentration with respect to each parameter. These derivatives can be used to calculate the gradient.

Benchmarking has shown forward sensitivity analysis to outperform finite differencing and non-gradient-based methods for small ODE systems  \cite{Raue2013}. The method has also been used to obtain reasonable fits for a library of benchmark problems \cite{Hass2018}, including models relevant to immune cell signaling \cite{Raia2011,Crauste2017}.
Most problems in this library featured systems of 5--30 ODEs. However, the forward sensitivity method generates an ODE system of size equal to the number of original equations times number of parameters \cite{Sengupta2014}. Stiff numerical integrators (e.g., CVODE \cite{Hindmarsh2005}) have a computational complexity limited by that of matrix multiplication \cite{Frohlich2019}, which is $n^3$ with a naive implementation and roughly $n^{2.38}$ with the best known algorithm (for a system of $n$ equations). This complexity limits the capacity to solve very large ODE systems. For example, for many systems derived from rule-based models (hundreds of ODEs, augmented to thousands of sensitivity equations), the forward sensitivity approach becomes intractable. 

\textit{Adjoint sensitivity} analysis uses a more complex mathematical framework to reduce the problem to the original integration combined with the (backward) integration of a newly derived adjoint system. The adjoint system has size equal to the number of parameters, making for a considerably smaller integration problem than with forward sensitivity analysis \cite{Sengupta2014}. The specific adjoint problem to be solved depends on the formulation of the optimization problem. One common case that has been implemented is minimization of a sum of squares objective function derived from time-series data. Fr\"{o}hlich et al. \cite{Frohlich2018} demonstrated that this implementation can be used to parameterize large biological models, whereas forward sensitivity analysis would be too computationally expensive. Adjoint sensitivity analysis is also promising for ODE systems derived from rule-based models, but currently lacks software support.

\textit{Automatic differentiation} (AD) has gained recent popularity given its applications to neural networks \cite{Baydin2018}. In principle, any algorithm can be represented as a computational graph (similar to a neural network) consisting of elementary computer operations. Derivatives of the algorithm outputs can then be calculated by propagating the derivatives of each operation in the graph via the chain rule. Although no biology-specific tools support AD, it is supported in the statistical modeling package Stan \cite{Carpenter2015,Carpenter2017}, where it can be applied to ODE models. Benchmarking of AD compared to other ODE sensitivity analysis methods suggests AD is efficient for small models, but scales poorly compared to adjoint sensitivity analysis \cite{Rackauckas2018}. It remains to be seen whether AD is computationally feasible for algorithms relevant for detailed biological models (i.e., algorithms for numerical integration of stiff and large initial value problems associated with ODE models).

A drawback of all forms of gradient-based algorithms is that each optimization run may only reach a local minimum or saddle point of the objective function. This limitation can be addressed by performing multiple replicates of optimization starting from different initial points (multistart optimization). Each additional replicate provides an additional opportunity to converge to the global minimum. 

\subsection*{Metaheuristic optimization}

Metaheuristic optimization algorithms \cite{Gandomi2013} are a family of methods that operate by repeated objective function evaluations, typically without the use of gradient information. Such algorithms aim to find a global (rather than local) optimum, and although they have no guarantee of good performance, they empirically provide good results in many use cases. Examples of such algorithms include evolutionary algorithms (e.g. differential evolution \cite{Storn1997} and scatter search \cite{Glover2000}), particle swarm optimization \cite{Eberhart1995}, and simulated annealing \cite{Kirkpatrick1983}. A feature of many but not all of these algorithms is the maintenance of a population of good parameter sets, which are used to generate new trial parameter sets. Many modern descriptions of population-based metaheuristic algorithms (e.g \cite{Penas2015a,Moraes2015,Penas2017}) allow for parallelized function evaluations within a single run of the algorithm, which enables these algorithms to take advantage of high-performance computing resources. 

Note that the parallelization of these algorithms is not simply from performing multiple fitting replicates (which can be trivially done for any algorithm); evaluations are parallelized within each iteration of the algorithm. Some metaheuristic algorithms (e.g., \cite{Moraes2015}) are asynchronous. Such algorithms improve load balancing by running simulations on all available cores at all times (cores are never left idle). In contrast, synchronous algorithms require all simulations of one iteration to complete before any core can move on to the next iteration. Both asynchronous and synchronous parallelized algorithms can use multiple cores to lower the total wall time required for fitting.  In contrast, multistart optimization (commonly used for gradient-based algorithms) requires the same wall time regardless of the number of cores used; it only increases the chance that some replicate finds a global optimum, as noted earlier. 

Metaheuristic optimization algorithms are useful for a range of problems for which gradient-based methods are not feasible. Such algorithms are implemented in PyBioNetFit and have been demonstrated on a library of problems \cite{Mitra2019a} including large rule-based models and stochastic models. A notable example problem features a rule-based model of TCR signal initiation simulated by network-free simulation \cite{Chylek2014a}.

Hybrid methods are available that incorporate both metaheuristic and gradient-based optimization. For example, many descriptions of scatter search (e.g., \cite{Egea2009}) include gradient-based local refinement of solutions found by the metaheuristic method. Such an algorithm outperformed both pure gradient-based and pure metaheuristic algorithms on a benchmark library \cite{Villaverde2019} featuring medium to large ODE models (tens to hundreds of ODEs and parameters). Local refinement of solutions can also be performed with gradient-free methods such as the simplex algorithm \cite{Nelder1965}, which can be parallelized \cite{Lee2007}. 

\section*{Parameter estimation using qualitative data}

In the above discussion, it was assumed that an objective function was derived from quantitative data. Recent methodological developments allow non-numerical, qualitative data to be leveraged in parameterization. These advances are notable because they allow new types of data, which may be easier to generate or already available in the literature, to be used in parameterization. 

An early example of using qualitative data is the work of Tyson and co-workers on the cell cycle \cite{Chen2000,Chen2004}. Successive versions of a model for cell cycle control were parameterized by hand-tuning, and in one case refined by an automated method \cite{Oguz2013}. Automated parameterization using qualitative data was also performed by Pargett et al. \cite{Pargett2013,Pargett2014}

In related, more recent work \cite{Mitra2018a}, qualitative data were formalized as inequality constraints imposed on the outputs of a model. We note that these inequality constraints on \textit{model outputs} differ from box constraints on parameter values, which are used in many parameterization problems. The inequalities were incorporated into the objective function (which can also include quantitative data) as static penalty functions \cite{Smith1997}. A static penalty function takes a value of zero when an inequality constraint is satisfied and a value proportional to the extent of constraint violation when a constraint is violated (and thus is shaped like the relu function used in machine learning). This method is available for general use in PyBioNetFit \cite{Mitra2019a}. 

\begin{figure}[tb!]
\centering\includegraphics{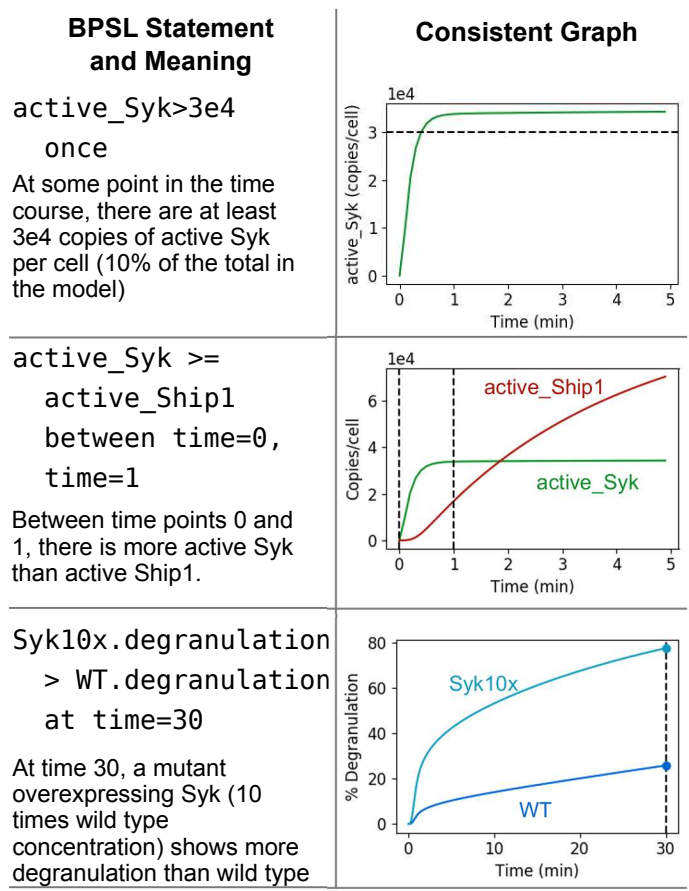}
\caption{Illustration of three statements in BPSL about a model of IgE receptor signaling. This model is adapted from ref. \cite{Harmon2017} using the published parameterization. The published model is consistent with all three of these BPSL statements}
\label{fig:qualtable}
\end{figure}

PyBioNetFit introduces the Biological Property Specification Language (BPSL) as a means to declare inequality constraints to be imposed on outputs of a model. BPSL is designed for the definition of qualitative properties of time courses or dose-response curves that might be observed experimentally. In particular, BPSL has \textit{enforcement keywords}, \texttt{always}, \texttt{once}, \texttt{at}, and \texttt{between}, which are used to declare where in a time course or dose-response curve an inequality should be enforced. For example, \texttt{always} indicates an inequality should be enforced at all points, and \texttt{at} indicates an inequality should be enforced at one specific value of the independent variable. BPSL also supports case-control comparisons, such as differences between mutant and wild type. Figure \ref{fig:qualtable} illustrates example BPSL statements applicable to a model of Fc$\epsilon$RI signaling. 

To our best knowledge, BPSL is the first language designed specifically for the definition of qualitative biological data. At present, the main use case is to configure parameterization in PyBioNetFit. More generally, BPSL can be seen as a knowledge engineering tool for formalizing qualitative information about the behavior of a biological system. This type of formalization has other applications, such as verifying that a given model agrees with known system properties (i.e., model checking) and for choosing perturbations of a system to achieve a desired set of properties (i.e., design). We hope that future development of other software tools can provide support for BPSL. 

\section*{Methods for Uncertainty Quantification}

While the above methods are useful for obtaining a parameterized model consistent with data, one should also ask how well identified are the model parameters and how uncertainty in parameter estimates propagates to uncertainty in model predictions. Such analysis is especially important when considering high-dimensional parameter spaces with limited experimental data. In such a case, we cannot reasonably expect to identify every parameter, but remarkably, we can often identify only some of the parameters and still make reliable predictions, as is the case for the model of ref. \cite{Harmon2017}.

\subsection*{Profile likelihood}

Profile likelihood \cite{Kreutz2013} is a relatively inexpensive method to assess the identifiability of model parameters. In this method, one parameter of interest is scanned over a series of fixed values. At each fixed parameter value considered, optimization of the objective function is repeated, allowing the values of all other free parameters to vary. Then the fixed parameter value is plotted against the minimum objective function value achieved in optimization. A smaller objective function value indicates a more likely value for the parameter. Prediction uncertainty can be calculated by an analogous approach \cite{Kreutz2012}. For ODE models, methods are available to calculate profiles by numerical integration instead of repeated optimization \cite{Hass2016,Kaschek2019}.

Profile likelihood has an underlying assumption that the objective function is related to the \textit{likelihood} of the parameters, that is, the probability of generating the experimental data given the chosen parameter set. In this framework, the chi-squared objective function corresponds to the negative log likelihood, under the assumption that the measurement errors were drawn from independent Gaussian distributions. 

Profile likelihood is useful for efficiently quantifying the identifiability of individual parameters and model predictions. It can be applied to high-dimensional parameter spaces for which other methods are not feasible. However, profile likelihood does not provide information on the relationships between parameters. For example, if a ratio of two parameters is identifiable but neither parameter is identifiable individually, this analysis will simply show both to be unidentifiable. 

\subsection*{Bootstrapping}

Bootstrapping is a method that performs uncertainty quantification by resampling data. 

In bootstrapping, the available dataset is first resampled. Multiple resampling methods are available \cite{Givens2013}, one example being to choose $n$ points \textit{with replacement} from an original data set of $n$ points \cite{Press2007}. Then the optimization algorithm is repeated on the resampled data, and the best fit is saved as a bootstrapped parameter set. The procedure is repeated many times to generate a desired number of bootstrapped parameter sets, which are examined to determine confidence intervals for each parameter.

The idea is that resampling the data is roughly equivalent to repeating the experiment. Thus resampling followed by refitting gives what can be thought of as a potential result if both the experiment and optimization were repeated. In the case of a perfect algorithm that always finds a unique global optimum, the result would depend only on the data. However, if optimization is biased (e.g., toward a local minimum near the starting point of optimization), this bias would appear in the bootstrap estimates of confidence intervals. 

\subsection*{Bayesian methods}

In Bayesian statistics, model parameters are taken to be random variables with unknown probability distributions. In this framework, uncertainty quantification is performed by finding the (multivariate) probability distribution of the parameters $\mathbf{\Theta}$ given the experimental data $\mathbf{y}$, $P(\mathbf{\Theta}|\mathbf{y})$. This distribution is proportional to $P(\mathbf{y}|\mathbf{\Theta})P(\mathbf{\Theta})$.  $P(\mathbf{y}|\mathbf{\Theta})$ can be calculated using a likelihood function (e.g., the chi-squared function is equal to $-\log P(\mathbf{y}|\mathbf{\Theta})$ plus a constant) and  $P(\mathbf{\Theta})$ is a user-defined prior. The distribution $P(\mathbf{\Theta}|\mathbf{y})$ is estimated using a sampling algorithm. A simple example is the Metropolis-Hastings (MH) Markov Chain Monte Carlo (MCMC) algorithm \cite{Chib1995}. MH-MCMC performs well in simple cases, but its efficiency declines for multimodal distributions \cite{Earl2005} and high-dimensional parameter spaces \cite{Betancourt2017}. The Gelman-Rubin statistic \cite{Gelman1992}, or a more recent improvement on it \cite{Vehtari2019}, can be used to determine when a sufficient number of MCMC samples have been collected.

More sophisticated sampling algorithms are available for cases where MH-MCMC is inefficient. Parallel tempering \cite{Earl2005} is a MCMC algorithm designed to improve sampling of multimodal distributions. If gradient information is available for the problem at hand, Hamiltonian Monte Carlo (HMC) \cite{Betancourt2017} is possible. By utilizing gradient information to make moves non-randomly, HMC can be more efficient than MH in high-dimensional parameter spaces \cite{Neal2012}. The No U-turn Sampler (NUTS) \cite{Hoffman2011} is a particularly useful version of HMC because with this method, algorithmic parameters are selected automatically. HMC can also be improved by choosing a problem-specific distance metric to make sampling more efficient \cite{Bales2019}.

Bayesian uncertainty quantification tends to be the most computationally intensive of the methods discussed here, but also provides the most complete picture of parametric uncertainty. The algorithm to calculate the multidimensional posterior probability distribution is unbiased by design (aside from the initial choice of a likelihood function and prior). The resulting distribution can be used to determine a confidence interval for each parameter (by examining the marginal distribution of the parameter) and to assess correlations between parameters. In addition, it is straightforward to quantify prediction uncertainty of the model by examining simulation outputs for sampled parameter sets. We recommend using Bayesian methods whenever computationally feasible. Such analysis was possible for a rule-derived ODE model of Fc$\epsilon$RI signaling \cite{Harmon2017} with 16 parameters and 23 equations, but is expected to be more challenging for higher dimensional parameter spaces. 

\section*{Software tools}

Most methods we have described have been implemented in recently developed, general-purpose software tools. Although many optimization tools are available, we limit our discussion here to tools that provide direct support for models supplied in standard formats (BNGL or SBML). Such tools are valuable because they enable parameterization and uncertainty quantification without the need for problem-specific coding. Such a workflow helps promote reproducible modeling \cite{Medley2016,Waltemath2016}, and allows modelers to focus on model analysis rather than debugging code. 

Table \ref{table:software} summarizes four major software tools that fit this use case: PyBioNetFit \cite{Mitra2019a}, COPASI \cite{Hoops2006,Bergmann2017}, Data2Dynamics (D2D) \cite{Raue2015}, and AMICI \cite{Frohlich2017}. All four of these tools remain under active development. An additional tool of note is PySB \cite{Lopez2013}, which has support for BNGL and SBML. PySB does not itself support parameterization or uncertainty quantification, but could be used in a Python program with other packages providing this functionality, such as Scipy (\url{https://www.scipy.org/}), MEIGO \cite{Egea2014}, or BayesSB \cite{Eydgahi2013}. Low-level parameterization packages are also available in R (e.g., MEIGO \cite{Egea2014} and dMod \cite{Kaschek2019}) and MATLAB (e.g., PESTO \cite{Stapor2018}).

The four tools considered in Table \ref{table:software} have different strengths, suitable for different applications. PyBioNetFit is unique in its support for BioNetGen models and simulators, and for including parallelization within its algorithms, making its metaheuristics more efficient than those of other tools, provided that they are run on a cluster or multi-core workstation. COPASI is notable for its ease of installation and user interface while providing many comparable features to other tools. D2D provides forward sensitivity analysis and a MATLAB interface. AMICI is the only tool supporting adjoint sensitivity analysis, but has a more difficult installation procedure than the other tools. 

\newcolumntype{W}{>{\raggedright\arraybackslash}m{4cm}}
\newcolumntype{V}{>{\centering\arraybackslash}m{2.2cm}}

\begin{table}[p!]
\small
\centering
\begin{tabular}{l|c c c c}
\hline
& PyBioNetFit & COPASI & Data2Dynamics & AMICI\\
\hline
\hline
Installation$^\textrm{\footnotesize{a}}$ & \multicolumn{1}{V}{Python package} & \multicolumn{1}{V}{Downloadable application} & \multicolumn{1}{V}{MATLAB source code} & \multicolumn{1}{V}{C++ source code}\\
\hline
User Interface$^\textrm{\footnotesize{b}}$ & Command-line & GUI & MATLAB & \multicolumn{1}{V}{MATLAB, Python, or C++}\\
\hline
\hline
BNGL support &\check&&&\\
\hline
SBML support &\check&\check&\check&\check \\
\hline
\hline
\multicolumn{1}{W|}{Gradient-based algorithms} && \footnotesize{c} &\check&\check \\
\hline
Adjoint sensitivity &&&&\check\\
\hline
Metaheuristic algorithms &\check&\check&\check&\\
\hline
Parallelized algorithms &\check& \footnotesize{d} & \footnotesize{d} & \footnotesize{d}\\ 
\hline
\multicolumn{1}{W|}{Optimization using qualitative data} &\check&&&\\
\hline
\hline
Numerical integration &\check&\check&\check&\check \\
\hline
Stochastic simulation &\check&\check&&\\
\hline
\hline
Profile likelihood &&\check&\check&\\
\hline
Bootstrapping &\check&&&\\
\hline
Bayesian methods &\check&&\check&\\
\hline
\end{tabular}
\caption{\normalsize{Summary of usage and features of four major software tools for parameterization and uncertainty quantification of models defined in BNGL or SBML.} \newline\newline
\footnotesize{a. PyBioNetFit is installed through the \texttt{pip} package manager. COPASI is a downloadable application that can be run without further configuration. Data2Dynamics is provided as MATLAB source code that can be run using commercial MATLAB software. AMICI is provided as C++ source code that must be compiled after performing machine-specific configuration of dependencies.} \newline
\footnotesize{b. PyBioNetFit is run on the command line using text files for configuration. COPASI has a GUI as well as a command line interface. Data2Dynamics provides functions that must be called through MATLAB code. AMICI provides functions that must be called through MATLAB, Python, or C++ code.} \newline
\footnotesize{c. COPASI's gradient-based algorithms use the finite difference approximation, making them less effective than those of other tools.} \newline
\footnotesize{d. These tools support parallelization of multiple independent fitting runs, but individual fitting runs cannot take advantage of parallelization.}}
\label{table:software}
\end{table}
\normalsize

\section*{Outlook}

Mathematical models will be increasingly important tools for understanding immunoreceptor signaling. New developments in software, coupled with increasing availability of computing resources, offer new opportunities for robust model parameterization. 

Parameterization using qualitative data is an exciting new direction that we hope to see explored in future work. At present, we have seen these methods applied to only a limited number of problems. The development of general-purpose software \cite{Mitra2019a} allows these methods to be applied broadly. At present the approach is limited to point estimation of parameters using static penalty functions. Uncertainty quantification would require software support for a statistical error model, which is not yet implemented, but is a promising future direction. 

Another promising direction given increased computational power is the parameterization of spatial models. One such example is a spatial model for receptor tyrosine kinase signaling \cite{Kerketta2016}, which was parameterized using problem-specific code. At present, multiple spatial simulators designed for biological applications are well-developed \cite{Andrews2010,Kerr2008} and in the future could be integrated with general-purpose parameterization tools.

These opportunities will enable the development of more detailed models supported by data, providing new means of studying cellular signaling processes. 

\section*{Acknowledgments}

We are grateful for support through grant R01GM111510 from the National Institute of General Medical Sciences (NIGMS) of the National Institutes of Health (NIH).

\section*{Declaration of Interest}

Declarations of interest: none.

\section*{Annotated References}

** \cite{Mitra2019a} The authors present PyBioNetFit, software for parameterization and uncertainty quantification of BNGL and SBML models, including parameterization using qualitative data. The software is benchmarked on a library of problems that includes rule-based models, ODE models, and stochastic models. 

** \cite{Frohlich2017} The authors present AMICI and its approach of adjoint sensitivity analysis for model parameterization.

** \cite{Kerketta2016} The authors use single-particle tracking data to parameterize a spatial model of receptor tyrosine kinase signaling. 

** \cite{Raue2015} The authors present Data2Dynamics, a MATLAB-based tool for the parameterization of ODE models. 

** \cite{Mitra2018a} The authors demonstrate an approach for using qualitative data in model parameterization and use this approach to parameterize an ODE model of cell cycle control. The model has 26 equations and 153 parameters.

** \cite{Frohlich2018} The authors parameterize a model with 1,200 equations and 4,100 parameters using adjoint sensitivity analysis, improved with multistart parallelization and a sparse numerical integrator.

* \cite{Hass2018} The authors present a library of parameterization problems featuring ODE models. The problems are solved using the gradient-based optimization methods of Data2Dynamics.

* \cite{Harmon2017} Metaheuristic optimization and Bayesian uncertainty quantification are applied to a model of IgE-Fc$\epsilon$RI-mediated signaling in mast cells.

* \cite{Villaverde2019} Gradient-based, metaheuristic, and hybrid optimization algorithms are tested on a library of benchmark problems featuring medium to large ODE systems.

* \cite{Bergmann2017} This review highlights the major features of the software COPASI, including its capabilities for parameterization. 

\bibliographystyle{model1-num-names}
\bibliography{bibliography.bib}

\end{document}